\documentclass[twocolumn,superscriptaddress,9pts]{revtex4}

\usepackage[dvips]{epsfig}
\usepackage{amsfonts}
\usepackage{amsmath}
\usepackage{}

\newcommand{\Eqref}[1]{(\ref{#1})}
\newcommand{\ket}[1] {\mbox{$ \vert #1 \rangle $}}

\newcommand{\bra}[1] {\mbox{$ \langle #1 \vert $}}
\newcommand{\abs}[1] {\mbox{$ \vert #1 \vert $}}

\newcounter{subequation}[equation] \makeatletter
\expandafter\let\expandafter\reset@font\csname
reset@font\endcsname
\newenvironment{subeqnarray}
  {\arraycolsep1pt
    \def\@eqnnum\stepcounter##1{\stepcounter{subequation}{\reset@font\rm
      (\theequation\alph{subequation})}}\eqnarray}
  {\endeqnarray\stepcounter{equation}}
\makeatother

\newcommand{\ba}{\begin{eqnarray}}
\newcommand{\ea}{\end{eqnarray}}
\newcommand{\sba}{\begin{subeqnarray}}
\newcommand{\sea}{\end{subeqnarray}}

\begin{document}

\vskip 1truecm

 \title{Quantum correlations in inflationary spectra
 and violation of Bell inequalities}
 \author{David Campo}
 \affiliation{Department of Applied Mathematics, University of Waterloo,
 University Avenue, Waterloo, Ontario, N2L 3G1 Canada}
 \author{Renaud Parentani}
 \affiliation{Laboratoire de Physique Th\'{e}orique, CNRS UMR 8627,
 B\^atiment 210, Universit\'{e} Paris XI, 91405 Orsay Cedex,
 France}

 \begin{abstract}

 In spite of the macroscopic
 character of the fluctuation amplitudes,
 we show that
 the standard inflationary distribution of primordial density fluctuations
 still exhibits inherently quantum mechanical correlations
 (which cannot be mimicked by any classical stochastic ensemble).
 To this end, we propose a Gedanken experiment
 for which certain Bell inequalities are violated.
 We also compute the effect of decoherence and
 show that the violation persists provided that the
 decoherence lies below a certain non-vanishing
 threshold.
 Moreover, there exists a higher threshold above which
 no violation of any Bell inequalities
 can occur, so that the
 corresponding distributions can be interpreted
 as stochastic ensembles of classical fluctuations.
 \end{abstract}
 \maketitle

 Inflation implies that the primordial density fluctuations arise
 from the amplification of vacuum fluctuations \cite{inflation}.
 As a result, the initial vacuum state
 becomes a product of highly squeezed two-mode states
 \cite{Grishchuk90} characterized by the wave vectors $\bf k$ and
 $-\bf k$.
 This follows from the homogeneity of the
 state and the linearity of the mode equation.
 However the linear treatment is only approximate.
 Hence, when no longer neglecting
 the weak non-linearities of inflationary models
 \cite{Maldacena03}, and when restricting attention to a given
 two-mode sector, decoherence will unavoidably develop.
 Since the interactions preserve the homogeneity
 and are weak (so as to justify the use of a Hartree approximation)
 inflationary distributions 
 belong to the class of Gaussian homogeneous
 distributions obtained by slightly decohering the standard
 distribution obtained by neglecting
 non-linearities \cite{KiefPolStar00,CP2,CPfat}.

 Given the macroscopic character of the fluctuation amplitudes
 at the end of inflation,
 one might ask whether some quantum properties
 have been preserved by these distributions,
 and if so,
 how to distinguish those which have kept some 
 from those which have lost them.
 To show that the standard
 distribution has
 kept quantum properties and to
 sort out these two classes of
 distributions, 
 we shall analyze 
 the possibility of violating Bell inequalities \cite{cpdraft}.

  \section{The inflationary spectra}

 \subsection{The 'standard' 
 distribution}

 In inflationary models based on one inflaton field,
 one can express the dynamics of the
 linear perturbations
 around the background
 solution of Einstein equations
 in terms of a massless 
 scalar field minimally coupled
 to gravity, $\phi$. This 
 field is the
 gauge-invariant linear combination of the
 gravitational potential and the inflaton fluctuations
 which satisfies canonical commutation relations
 \cite{MukhaPhysRep}.
 We briefly outline how one obtains the
 two-mode squeezed states and the time-coherence of the modes
 \cite{Grishchuk90,CP2}.

 The background geometry is a
 Friedman-Robertson-Walker space-time, with line element
 \ba
   ds^2 =
   a^2(\eta) \left[ - d\eta^2 + \delta_{ij} dx^i dx^j\right]
   \, .
 \ea
 The spatial sections in
 the present Hubble volume are flat to an
 excellent precision
 if inflation has lasted enough to have increased the radius
 of curvature to super-horizon scales.
 The scale factor, as a function of the conformal time $\eta$
 is given by $a(\eta) = -1/H\eta$ in the
 inflationary period we approximate by a de Sitter space.
 At the reheating $\eta_r$, we paste this law
 to a radiation dominated phase
 characterized by $a(\eta) = (\eta - 2\eta_r)/(H\eta_r^2)$.
 We can then follow the time dependence of modes
 from the onset of inflation till horizon
 re-entry in the radiation dominated era.

 After expanding the action of matter and metric
 fluctuations up to quadratic order around the background
 solution,
 the Hamiltonian for each
 complex Fourier mode $\phi_{\bf k}$ is
 \ba \label{Hquad}
   H_ {\bf k} =  %
   \frac{1}{2} 
   \, \left[
   \vert  \partial_{\eta}  \phi_{\bf k} \vert^2
   + \left(k^2 - \frac{\partial_{\eta}^2 a}{a} \right) \,
  \vert   \phi_{\bf k} \vert^2
  \right]
   \, .
 \ea

 The state of $\phi_{\bf k}$ at the onset of inflation
 follows from the kinematics of the background
 \cite{RenaudCMB}:
 if some inhomogeneities are present at an early stage of inflation,
 they are diluted by the quasi-exponential
 expansion over proper lengths bigger than our causal patch, i.e.,
 to us, they are indistinguishable from the background. One is thus
 left with the vacuum.
 One can also reach this conclusion by considering
 the evolution of perturbations
 backward in time. Then, non-vacuum contributions
 (classical inhomogeneities as well as quantum excitations of the
 field) to the energy-momentum tensor blow-up thereby
 violating the consistency condition of no-large backreaction.
 In conclusion,
 all the modes observable today were in their ground
 state about $70$ e-folds before the end of inflation
 (the minimal duration of inflation to include
 today's horizon inside a causal patch).

 This ground state is  unambiguously defined for the relevant
 modes because, at that time, their wavelength was
 much shorter than the Hubble radius (i.e.
 $k^2 \gg \partial_{\eta}^2 a/a$
 in Eq. \Eqref{Hquad}).
 One therefore works with the vacuum
 defined in the asymptotic past by
 positive frequency modes
 \ba \label{positivefreq}
   \lim\limits_{\eta \rightarrow -\infty}
   \left[ i\partial_{\eta} - k  \right]\phi_{k}^{in}(\eta) = 0
   \, ,
 \ea
 where $\phi_{k}^{in}(\eta)$ are solution of the mode equation:
 \ba \label{modeeq}
   &&\partial_{\eta}^2 \hat \phi_{\bf k} +
   \left(k^2 - \frac{\partial_{\eta}^2 a}{a} \right) \,
   \hat \phi_{\bf k} = 0 \, , \nonumber \\
   &&\hat \phi_{\bf k}(\eta) =
   \hat a_{\bf k}^{in}\phi_{k}^{in}(\eta) +
   \hat a_{-{\bf k}}^{in\, \dagger} \phi_{k}^{in\,*}(\eta) \, .
 \ea

 The time-dependent term $\partial_{\eta}^2 a/a$
 mixes the positive ${\bf k} $ and negative frequency ${\bf -k}$
 modes. Therefore
 the vacuum unitarily evolves into an entangled state:
 a two-mode squeezed state.
 Confusion about the origin of this entanglement seems to persist:
 entanglement is {\it not} a consequence of the reality (self-adjointness)
 of $\hat \phi(\eta, {\bf x})$.
 Pair production of charged particles
 (in an electric field) also leads to entangled states.
 Two-mode entanglement is due to the mixing of positive and negative
 frequencies and to the homogeneity of the background.


 At the end of inflation, $\partial_{\eta}^2 a $ vanishes
 and the amplification stops. During the radiation dominated era,
 in the absence of iso-curvature modes, the fluctuations
 of physical fields
 (those of the gravitational potential and of
 cold dark matter and radiation density fluctuations)
 are all linearity related to the values of $\phi$ and
 $\partial_{\eta}\phi$ evaluated
 at the end of inflation. This is true both classically and
 quantum mechanically because of the linearity of the transfer matrix.

 Instead of working with these physical fields, it is simpler
 to keep working with $\phi_{\bf k}$, since its time evolution is
 simply governed by $e^{\pm ik\eta}$.
 We shall use this behavior to further simplify the writings
 by introducing creation and destruction operators
 $\hat a_{\bf k}, \, \hat a_{\bf k}^{\dagger} $ defined
 in the adiabatic era by
 $\hat \phi_{\bf k} = (\hat a_{\bf k} e^{- ik\eta} + h.c.)/\sqrt{2 k}$.
 In terms of these operators, the `in` vacuum 
 (i.e. the state annihilated by the $\hat a^{in}_{\bf k}$ of
 Eq. (\ref{modeeq})) reads \cite{Grishchuk90}
 \ba \label{inoutvac}
   \ket{0,\, in} &=&
   \widetilde{\prod_{\bf k}}
      \left(1-\abs{z_k}^2 \right)^{1/2} \nonumber \\
      &\times& \exp \left( z_{k}\, \hat  a_{\bf k}^{ \dagger}  \,
      \hat a_{-\bf k}^{\dagger} \right) \,
      \ket{0,\, \bf k}\otimes \ket{0,\, -\bf k}
      \, .
 \ea
 The tilde tensorial product takes into account only half the
 modes. It must be introduced because the
 (squeezing) operator
 $\exp \left( z_{k}\, \hat  a_{\bold k}^{ \dagger}  \,
 \hat a_{-\bold k}^{\dagger} \right)$
 acts on pairs1 of modes.
 The complex number $z_{k}$ characterizes the
 two-mode states of wavenumber $k=\abs{\bf k}$.
 It is related to the following expectation values
 \ba \label{momenta}
   \langle \hat a_{\bf k}^{\dagger} \,   \hat  a_{\bf k'} \rangle
   = n_k  \, \delta^{3}({\bf k} - {\bf k}' ) \,
   \, , \quad
   \langle \hat a_{\bf k}  \, \hat a_{\bf k'} \rangle
   = c_k \, \delta^{3}({\bf k} + {\bf k}' )
   \, .
 \ea
 The mean occupation number is  $n_k = \abs{z_k}^2/(\abs{z_k}^2 - 1)$
 and the cross correlation term is $c_k = z_k/(\abs{z_k}^2 - 1)$.
 They obey $\abs{c_k}^2= n_k(n_k + 1)$.
 This is the highest value of $\abs{c_k}$ for a given $n_k$.
 It is obtained for the states which
 minimize Heisenberg uncertainty relations \cite{CPfat}.

 Asking that the spectrum
 describes fluctuations with r.m.s.
 amplitude $10^{-5}$
 for wavelength observable in the CMB,
 one gets the rough estimation $n_k \sim 10^{100}$.
 The phase of $z_k$ determines the temporal phase of the modes as they
 re-enter the Hubble radius \cite{CPfat}. Namely, the power spectrum
 of $\phi_{\bf k}$ at conformal time $\eta$ is given by a product of two modes
 proportional to $ e^{i k \eta} + z_k e^{- i k \eta}$. In
 inflationary models, $z_k$ is very close to $-1$, giving rise to
 $\sin(k \eta)$ functions, i.e. to 'growing modes'.


 \subsection{
 Partially decohered Gaussian distributions}

 The two-mode entanglement of squeezed states
 follows from the homogeneity of the
 vacuum and the linearity of the mode equation. However,
 as discussed in the introduction, the linear
 treatment is only approximate. When no longer neglecting
 the weak non-linearities of
 gravitation
 \cite{Maldacena03}, and when restricting attention to a given
 two-mode sector, some decoherence develops.
 This is very similar to the description of an experiment of
 diffraction of a slightly anharmonic cristal. The calculation of
 the reduced density matrix of the two modes under scrutiny is
 very difficult.
 But the weakness of the interactions guarantees that
 the modifications of the power spectrum are extremely small. It also
 guarantees that the reduced density matrix can still be
 accurately described by a Gaussian distribution. Moreover, since
 homogeneity is preserved, these distributions always factorize
 into products of uncorrelated two-mode sectors. In conclusion,
 inflationary distributions of primordial fluctuations belong to
 the class of Gaussian homogeneous distributions obtained by
 (slightly) decohering the
 standard distribution
 $\hat \rho_{in} = \ket{0, \, in}\bra{0, \, in}  $
 \cite{KiefPolStar00,CPfat}.

 These are described by products of two-mode density matrices
 $\hat \rho_2$ which are characterized by the expectation values
 \Eqref{momenta}, where $n_k$ and $\abs{c_k}$ are now
 {\it independent} parameters.
 We shall work at fixed $n_k$ (fixed power spectrum), and
 fixed $\arg(c_k)$
 (changing this phase does not change the entropy \cite{CPfat}),
 and let the norm $\abs{c_k}$ vary.
 A convenient parametrization is provided by $\delta_k$:
 \ba
   \abs{c_k}^2 = (n_k+1)(n_k-\delta_k) \, .
 \ea
 Squeezed states \Eqref{inoutvac} are maximally coherent and
 correspond to $\delta_k = 0$.
 The least coherent distribution,
 a product of two random density matrices, corresponds
 to $\delta_k = n_k$.
 Hence, the parameter $\delta_k$ controls
 the level of decoherence of the distribution,
 or equivalently the strength of
 the correlations between the ${\bf k}$ and the $-{\bf k}$ sectors.
 It also fixes the r.m.s. amplitude of the
 {\it decaying} mode, as seen
 by decomposing the field modes $\hat \phi_{\bf k}$
 in terms of
 the growing and the decaying modes:
 \ba
   \hat \phi_{\bf k}(\eta) =
   \hat g_{\bf k} \frac{\sin (k\eta)}{\sqrt{k}} +
   \hat d_{\bf k} \frac{\cos (k\eta)}{\sqrt{k}} \, .
 \ea
 When $\delta \ll n$,
 \ba
   \langle \hat g_{\bf k} \hat g_{\bf k}^{\dagger}
   \rangle &=& n + \frac{1}{2} - Re(c) =
   2n \left( 1 + O(\frac{\delta}{n})\right)
   \, , \nonumber \\
   \langle \hat d_{\bf k} \hat g_{\bf k}^{\dagger}
   \rangle &=& n + \frac{1}{2} + Re(c) =
   n^{1/4}\left(1 + O(\frac{\delta}{n})\right)
   \, , \nonumber \\
   \langle \hat d_{\bf k} \hat d_{\bf k}^{\dagger}
   \rangle &=&  Im(c) = \frac{\delta}{2} + O(n^{-1/2})
   \ll \langle \hat g_{\bf k} \hat g_{\bf k}^{\dagger}
   \rangle
   \, .
 \ea
 We have dropped the indexes $k$ since each two-mode can be analyzed separately
 (unless one is considering localized wave packets).

 We emphasize that the increase of entropy
 from the initial pure vacuum state
 to a statistical mixture follows, as usual, from the
 neglect of correlations, see footnote p. $188$
 in \cite{Gottfried1}.
 In other words, ignoring the correlations between several two-modes
 is the coarse graining we adopt.
 We also notice that this Gaussian ansatz is a first order
 (Hartree) approximation.
 With the future experiments,
 deviations from Gaussianity will hopefully be measurable
 and will become a powerful tool
 to raise the degeneracy between various
 models for the primordial universe \cite{RiottoNGreview}.

 \section{Clauser-Horne inequalities for
 inflationary spectra}

 \subsection{Local Hidden Variable completions of Quantum
 Mechanics and the Clauser-Horne inequality}

 Bell inequalities and their generalizations form a set of
 constrains on the statistics of outcomes of measurements
 when one demands that these outcomes
 are compatible with the classical concept of
 locality \cite{Gottfried2}.

 Consider a system made of two subsystems I and II
 and let the two parts interact.
 In general they will become correlated.
 Suppose that after 'turning off' the interactions,
 the system is in a pure state of the form
 \ba \label{entangledstate}
    \ket{\chi} = \sum_{n} A_n \ket{\phi}^{I}_n \otimes
    \ket{\psi}^{II}_n \, ,
 \ea
 where there is more than one term in the sum,
 where $\ket{\phi}^{I}_n$ and $\ket{\psi}^{II}_n$ are two
 sets of orthonormal states in the Hilbert spaces
 ${\cal H}^I$ and  ${\cal H}^{II}$ of I and II
 respectively, and where $P_n= |A_n|^2 < 1$
  are the probabilities for the
 $n$-st outcome to be realized
 such that $\sum_n P_n = 1$.
 The states \Eqref{entangledstate}
 are a special case of entangled states,
 see Section III.C for the definition of entanglement
 for statistical mixtures. (Notice that each two-mode squeezed state
 in Eq. \Eqref{inoutvac} is an entangled state.)
 Being orthonormal, the $\ket{\phi}^{I}_n$ and
 $\ket{\psi}^{II}_n$ can be seen as
 eigenstates of observables $\hat \Phi^I \otimes {\bf 1}^{II}$ and
 ${\bf 1}^{I} \otimes \hat \Psi^{II}$ acting non-trivially
 on ${\cal H}^I$ and  ${\cal H}^{II}$ respectively.

 After having let the two subsystems interact,
 we separate them to two distant places
 where two measurement separated by a space-like
 interval will be performed.
 We suppose that we have a ensemble of copies of the same
 system all prepared in state $\ket{\chi}$
 in order to perform different measurements and to accumulate
 statistical data.
 The space-like character of the interval
 guarantees that the state of one subsystem cannot
 be affected by any measurement performed on the other
 subsystem.
 Once the apparatus I has measured
 the observable $\hat \Phi^I$ on system I, we
 can predict with probability $1$ (as checked by
 reproducing the experiment in the same conditions a large
 number of times and bringing the data together)
 the outcome of the measurement of
 ${\bf 1}^{I} \otimes \hat \Psi^{II}$,
 irrespectively of the values of the probabilities $P_n$
 and despite the space-like separation of the events of the
 measurements on subsystems I and II (supposing
 we have the {\it a priori} knowledge that the system is
 in the entangled state $\ket{\chi}$,
 as we do in inflationary cosmology, see eq. (6)).

 In order to explain these correlations
 one might be tempted
 to consider that they result from
 unknown common properties shared by the two
 subsystems, and whose values were assigned while I and II were
 still interacting.
 These properties, to which we do not have access by
 the experiments, are called
 Hidden Variables.
 The Hidden Variable Programm consists in
 establishing the conditions
 which have to be met.
 In particular, Bell's theorem and its generalizations
 \cite{BellWigner,MerminReview,Gottfried2} show that
 Quantum Mechanics cannot be embedded
 in a {\it Local} Hidden Variable Theory.
 In this paper,
 we will consider a
 particular type of inequalities
 called the Clauser-Horne (CH) inequalities  \cite{CH,Gottfried2}.

 \subsubsection{Local Hidden Variable Theory}

 A Local Hidden Variable Theory (LHVT) is the specification of \\
 i) the conditional probability $p_{\Phi}[\phi;\, \lambda]$
 (resp. $p_{\Psi}[\psi;\, \lambda]$) that the measurement of
 $\hat \Phi^{I}$ (resp. $\hat \Psi^{II}$)
 gives the eigenvalue $\phi$ (resp. $\psi$),
 given a value $\lambda$ of the hidden variables, \\
 ii) a probability measure $p(\lambda) d\lambda$.

 The requirement of locality forces that \\
 1) the conditional probabilities $p_{\Phi}$
 (resp. $p_{\Psi}$) is independent of
 $\hat \Psi^{II}$ (resp. $\hat \Phi^I$), \\
 2) the probability distribution $p(\lambda)$ is independent of
 both $\hat \Phi^I$ and $\hat \Psi^{II}$, \\
 3) given a value $\lambda$ of the hidden variables, the outcome
 of the measurements of $\hat \Phi^I$ and $\hat \Psi^{II}$ are
 statistically independent, i.e.
 \ba \label{locality}
   p_{\Phi \, \Psi}(\phi, \psi ; \, \lambda) =
   p_{\Phi}[\phi;\, \lambda] \times p_{\Psi}[\psi;\, \lambda]
   \, .
 \ea

 A given quantum state
 $\hat \rho \in {\cal H}^{I} \otimes {\cal H}^{II}$
 admits a LHVT if, and only if,
 for any pair of observables with spectral decompositions
 $\hat \Phi^{I} = \sum_{n} \phi_n \hat \Pi_n^{\Phi}$ and
 $\hat \Psi^{II} = \sum_{n} \psi_n \hat \Pi_n^{\Psi}$,
 the statistical predictions of the model reproduce 
 those of Quantum
 Mechanics, i.e. the joint probability is
 \ba \label{clcorr}
   P_{\Phi\, \Psi}(\phi,\psi) &=& \int\!\!d\lambda\, p(\lambda) \,
   p_{\Phi \, \Psi}(\phi, \psi ; \, \lambda)
   \nonumber \\
   &=& {\rm Tr}\left(
   \hat \rho \hat \Pi_n^{\Phi} \hat \Pi_n^{\Psi}\right) \, .
 \ea

 \subsubsection{Clauser-Horne inequalities}

 The Clauser-Horne inequality is a condition verified by
 the joint probabilities $P_{\Phi\, \Psi}(\phi,\psi)$ if
 Quantum Mechanics
 can be embeded in a LHVT. They
 follow from the algebraic inequality satisfied by any set of
 four real numbers $(x,x',y,y')$, all lying in the interval
 $\left[0,\, 1 \right]$, see \cite{CH} for a proof:
 \ba \label{algebra}
   xy + xy' + x'y - x'y' - x - y \leq 0 \, .
 \ea
 Identifying $x=p_{\Psi}[\psi; \lambda], \, x'=p_{\Psi}[\psi; \lambda]
 ,\, y = p_{\Psi}[\psi; \lambda], \,
 y' = p_{\Psi}[\psi'; \lambda]$, and averaging
 over $\lambda$ one gets
 \ba \label{CHineq}
   \left[P_{\Phi\, \Psi}(\phi,\psi) + P_{\Phi\, \Psi}(\phi,\psi')
   + P_{\Phi\, \Psi}(\phi',\psi) \right.
   \qquad &&\nonumber \\
   - \left. P_{\Phi\, \Psi}(\phi',\psi')
   \right]
   \times
   \left[P_{\Phi}(\phi) + P_{\Psi}(\psi)\right]^{-1}
   \leq 1 \, ,
 \ea
 where $P_{\Phi},\, P_{\Psi}$ are the marginal probabilities.

 \subsection{A set of observables for Gaussian states}

 In searching for a violation of
 Bell inequalities in inflationary spectra, one encounters an unexpected
 difficulty because we are dealing with Gaussian states.
 Indeed, observables which are polynomes of the
 field amplitude and its conjugate momentum  cannot violate Bell
 inequalities.
 The reason is that
 the Wigner representation of Gaussian density matrices is
 positive.
 Hence, it can be used as a probability distribution function to build an
 LHV description of the expectation values \cite{BellWigner}.

 We must therefore consider observables which do not have a
 direct classical counterpart.
 Nonetheless the operators we shall consider have a clear and
 simple meaning. They are projectors on a given pair of coherent states
 $\ket{v, {\bf k}}$ and $\ket{w, {\bf -k}}$
 \ba  \label{proj}
   \hat \Pi(v,w) = \ket{v, {\bf k}}\bra{v, {\bf k}} \otimes
   \ket{w, {\bf -k}}\bra{w, -{\bf k}} \, .
 \ea
 The coherent states obey, by definition,
 $\hat a_{\bf k} \ket{v, {\bf k}} = v \ket{v, {\bf k}}$
 and
 $\hat a_{\bf - k} \ket{w, {\bf -k}} = w \ket{w, {\bf -k}} $.

 We want make a pose and
 advertise the pertinent role of coherent states
 in inflationary cosmology.
 Firstly, for quadratic Hamiltonians, they are minimal uncertainty
 Gaussian states which are stable in time.
 They are thus the quantum
 counterparts of classical points in phase space. More generally,
 one expects that they are still
 stable as long as the secular effects of
 non-linearities (changes in the width of coherent states,
 formation of small scale structures -whirls and tendrils- of the
 classical trajectories \cite{Berry}) remain small.
 Such is the case in cosmology as long as the the linear regime
 (and therefore the Gaussian approximation) are good
 approximations.

 Secondly, for quadratic
 systems weakly interacting with a reservoir of modes
 (here the coupling is through the non-linear gravitational
 terms),
 there exists a finite time after which
 the system reaches
 a state which is statistical mixture of coherent states
 (more generally, of minimal uncertainty Gaussian states)
 \cite{PazZurek,Eisert} \footnote{A word
 of caution is in order. This result is
 demonstrated for finite temperatures only, and might not be valid
 for $T = 0$.
 Nevertheless this result should remain true since the
 amount of decoherence necessary to kill the quantum correlations
 corresponds to add incoherently $1$ photon
 to each mode, see Section III.C.
 That this is realized in inflationary cosmology remains to be seen.}.

 Coherent states can therefore be thought as providing a particular
 realization of
 the Gaussian ensemble describing the primordial
 fluctuations.
 The probability that the semi-classical values $v$ and $w$
 of the ${\bf k}$-th mode be realized is
 \ba \label{proba}
    Q(v,w; \delta) &=& {\rm Tr}[\hat \rho_2(\delta) \, \hat \Pi(v,w) ]
    \nonumber\\
    &=& Q_{v}(v) \times Q_{w|v}(w; \delta)
    \, , \nonumber \\
    &=& \frac{1}{n+1} \exp\left[ - \frac{\abs{v}^2}{(n+1)} \right]
    \nonumber \\
    &&\times
    \frac{1}{1+ \delta} \exp\left[ -
    \frac{\abs{w - \bar w(v)}^2}{1+\delta}  \right]
    \, . \quad
 \ea
 We have written this probability
 in an asymmetric form to make explicit the
 power spectrum ($=n+1$), and the much smaller width
 ($={1+\delta}$) governing the dispersion of the values of $w$
 around $\bar w(v) =  v^* c/(n+1)$.
 Hence, as long as $\delta \ll n$, the
 amplitude of the mode $-{\bf k}$ conditional to the detection of
 the mode ${\bf k}$ in the coherent state $\ket{v}$ is
 fixed by that $v$.
 We see explicitly that it is $\delta$, the
 decoherence level,
 which governs  the strength of the correlations between the
 $\bf k$ and the $-{\bf k}$ sectors.

 Thirdly, given a realization, one can calculate
 conditional correlations. The latter
 have specific spatial properties best
 revealed by wave-packets.
 Although their Fourier content and their localization depends on
 the chosen wave-packet,
 their space-time structure is
 uniquely determined by the frequency mixing.
 More precisely, given the ensemble of realizations
 selected by $ \hat \Pi_{w.p.}$ the projector on some wave packet,
 the conditional field amplitude is
 (see \cite{CP2} for more details)
 \ba
   \langle \hat \phi(\eta,{\bf x}) \rangle =
   \frac{\langle \hat \Pi_{w.p.} \, \phi(\eta,{\bf x}) \,
   \hat \Pi_{w.p.} \rangle }{\langle \hat \Pi_{w.p.} \rangle }
   = \bar \phi_{R} +  \bar \phi_{L}
   \, ,
 \ea
 where
 \ba
   \bar \phi_{R}(\eta, {\bf x}) &=&
     \int\!\! \frac{\widetilde{d^3 k}}{(2\pi)^{3/2}} \, \left(
     v_{\bold k} \, 
     {e^{i \bold k \bold x}}
     \phi_{k}^{out}(\eta) + c.c.  \right)
     \, , \nonumber \\
    \bar  \phi_{L}(\eta, {\bf x}) &=&
    \int\!\!\frac{\widetilde{d^3 k}}{(2\pi)^{3/2}} 
    \,
     \left( \bar w_{k}(v_{\bf k}) \, 
    {e^{-i \bold k \bold x}}
     \phi_{k}^{out}(\eta) + c.c. \right) \nonumber
  \, .
 \ea
 One sees that the centers
 of these wave-packets propagate in opposite directions
 on the future light cone from the event of creation
 at the reheating. Hence, on a given time slice, e.g. the Last
 Scattering Surface (in the matter dominated era),
 they are separated by more that twice the Hubble radius at that
 time.

 The important conclusion for Bell inequalities is that
 the measurement \Eqref{proj}
 (more precisely, its generalization which contains a product over ${\bf k}$
 so as to describe a localized wave packet)
 can be performed on subsystems which
 are causally disconnected.
 Violations of Bell inequalities with 
 such observables
 were first considered by W\`{o}dkiewicz  {\it et al} in the
 context of Quantum Optics \cite{Wodkievicz98}.

 \subsection{Separable states}

 A two-mode state is called separable,
 or classically correlated, if and only
 if it can be written as a convex sum of products of
 one-mode density matrices \cite{Werner}. Otherwise, the state is
 called entangled.
 Restricting to the class of homogeneous Gaussian states
 \Eqref{momenta}, separable states are of the form
 \cite{SimonCirac}
 \ba \label{separable}
    \hat \rho_2^{\rm sep.}(\delta) =
    \int\!\!\frac{d^2v}{\pi} \frac{d^2w}{\pi} \,
    P(v,w;\delta) \, \hat \Pi(v,w) \, ,
 \ea
 where $P$ is a Gaussian function.
 For homogeneous distributions, it is given by \cite{CPfat}
 \ba
   P(v,w;\delta) &= &
   \frac{1}{\Delta'} \, 
   e^{-|v|^2 / n}
   \times
   \exp \left[ - \frac{\abs{w - (c v^*/n)}^2}{\Delta'/n }
   \right] \, , \nonumber
 \ea
 where $\Delta' = n^{2} - \abs{c}^2$ must be positive.
 This implies $\abs{c} \leq n$, or $\delta \geq n/(n+1) > 1$.
 In other words, all the states such that
 $n < \abs{c} \leq \sqrt{n(n+1)}$ are entangled.

 The physical meaning of separable states is revealed by
 the fact that they can be prepared by
 a classical protocol:
 a random generator
 produces the four real numbers encoded in $(v,w)$
 with probability $P$. The result of each drawn is send by
 classical communication channels to
 two distant observers performing separate
 measurements on the subsystems
 $\bf k$ and $-{\bf k}$ respectively so as to
 prepare them
 into the two-mode coherent state
 $\ket{v, {\bf k}}\ket{w, - {\bf k}}$.
 By construction, the statistical properties of separable states
 can be interpreted classically.
 In particular, separable states cannot violate Bell inequalities
 as demonstrated in \cite{Werner}. We will use them to derive the
 CH inequalities.

 When applied to separable Gaussian states \Eqref{separable},
 the probability \Eqref{proba} is
 \ba
  && Q^{\rm sep.}(v,w;\delta) =
  \int\!\!\frac{d^2\alpha}{\pi}\frac{d^2\beta}{\pi} \,
   P(\alpha,\beta;\delta) \, p(v|\alpha) \, p(w|\beta) \, ,
 \nonumber
 \ea
 where $p(v|\alpha) = \abs{\langle v \vert \alpha \rangle}^2
 = e^{-\abs{v-\alpha}^2} \leq 1$. This has the structure of
 Eqs. \Eqref{locality} and \Eqref{clcorr}, where
 $(\alpha,\, \beta)$ play the role of the ''hidden variable''
 $\lambda$, and $P(\alpha,\beta;\delta)$ the role of the probability
 density $\rho(\lambda)$.
 Then Eq. \Eqref{CHineq} gives
 \ba \label{Bell}
   {\cal C}(v,w;\delta) &=& [Q(0,0;\delta) + Q(v,0;\delta)
   \\
   &+&Q(0,w;\delta) - Q(v,w;\delta)]
   \times \frac{n+1}{2}
   \leq
   1 \, .
  \nonumber
 \ea

  \section{Violation as a function of the level of coherence}

 We now have all the elements to search for
 combinations of $v$ and $w$ which maximize ${\cal C}$ of Eq. \Eqref{Bell}.
 The maximum is reached for $\arg(c^* v w) = \pi$ and $\abs{v}= \abs{w}$.
 To make contact with \cite{Wodkievicz98}, we choose the arbitrary phase
 $2 \arg(v)= \arg(c)$ to get $w=-v$.
 The maximum of this function is reached for
 \ba
   \frac{\abs{v_M(\delta)}^2}{1+ \delta} &=&
   \frac{\ln \left[ 1 + \sqrt{\frac{n-\delta}{n+1}} \right]}{
   1 + 2 \sqrt{\frac{n-\delta}{n+1}}} \nonumber \\
   &=& \frac{\ln 2}{3}
    \left[ 1 + O\left(\frac{1 + \delta}{n}\right)
   \right] \, ,
 \ea
 and its value is
 \ba \label{vi}
  {\cal C}_M(\delta) &=& \frac{1}{2(1+\delta)}
  \left[ 1 + \frac{3}{2^{4/3}}
  +  O\left(\frac{1 + \delta}{n}\right)
  \right] \, . \quad
 \ea
 %
 %
 There is a violation even for macroscopic occupation numbers.
 This fact demonstrates that $n \gg 1$ is a necessary but
 {\it not} sufficient condition to have a
 classical 
 (stochastic) distribution.
 The latter are only 
 the separable ones for which
 $\delta \geq 1$.

 More precisely, the inequality \Eqref{Bell} is violated for
 \ba  \label{range}
   \delta < \frac{-1+3/2^{4/3}}{2} \simeq 0.095 \, ,
 \ea
 {\it irrespectively} of the value of $n$ when $n \gg 1$.
 Even though each probability in \Eqref{Bell}
 decreases like $1/n$,
 the range of $\delta$ is asymptotically constant in the regime relevant for
 inflationary cosmology.
 One also verifies that, as expected, the maximal violation is obtained for
 $\delta = 0$, i.e. for pure states.
 In Fig. 1
 we have plot ${\cal C}(v)$ for three values of $\delta$.

 \begin{figure}[h]
    \resizebox{7.5cm}{6.5cm}{\includegraphics{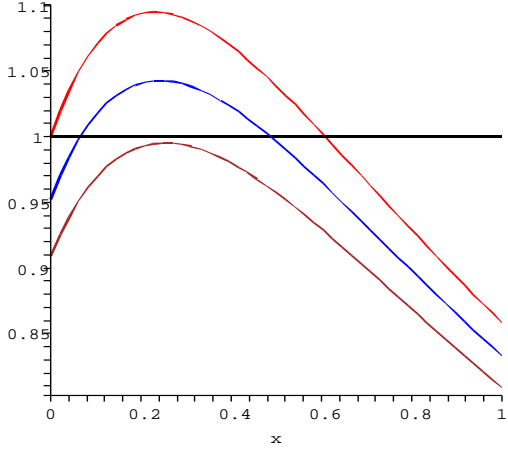}}
 \caption{{\it The loss of violation as decoherence increases.}
 The function is normalized to its maximal classical value: ${\cal C}=1$.
 The variable $x= \abs{v}^2$ gives the square amplitude of the field
 configurations.
 The occupation number is $n=100$, and the three values of $\delta$
 are $0$ (red, upper), the pure state, $0.05$ (blue, middle), and
 $0.1$ (brown, lower), the border regime.
 The function ${\cal C}(v, -v, \delta)$
 is asymptotically independent of $n$ in the large $n$ limit.
 }
 \end{figure}

 \section{Final remarks}

 First, there might
 be operators which violate Bell inequalities for
 $0.1 \leq \delta < 1$.
 However it is not guaranteed that these will have a simple
 interpretation in cosmological terms.

 Second, an observational verification
 of the violation would require several steps.
 First, one should be able to distinguish $\bf k$ from
 $- \bf k$ configurations,
 i.e. to have access to the primordial velocity field, see \cite{CP2}.
 Second, one should isolate rare
 realizations, far from the r.m.s. values,
 and specified with a precision given by
 the spread of the coherent states
 ($=1$) which is also much smaller than $n$.
 Given that  $n \sim 10^{100}$,
 an  observational verification
 seems therefore excluded.

 It is nevertheless conceptually important to
 realize that if 
 non-linearities, and hence decoherence, are weak enough so that
 $\delta < 1$, the distribution of primordial density fluctuations
 would have kept its quantum properties 
 in spite of the macroscopic character of the amplitudes and
 even though we cannot observationally verify it.

\end{document}